\documentclass[twocolumn,slac_two]{revtex4} 
\usepackage{latexsym}
\usepackage{graphicx}
\usepackage{fancyhdr}
\renewcommand{\bar}[1]{\overline{#1}}

\newcommand{\ket}[1]{\,\left|\,{#1}\right\rangle}
\renewcommand{\headrulewidth}{0pt}

\begin{document}

\vbox{\hbox to\paperwidth{\hfill SLAC-PUB-12125}
\hbox to\paperwidth{\hfill October 2006}}
\vspace{.5in}

\title{
Hadron Spectroscopy and Structure from AdS/CFT}

\author{Stanley J. Brodsky}

\affiliation{Stanford Linear Accelerator Center\\
Stanford University\\
Stanford, California 94309}

\begin{abstract}
The AdS/CFT correspondence between conformal field theory
and string states in an extended space-time  has provided new
insights into not only  hadron spectra, but also their light-front
wavefunctions. We show that there is an exact correspondence between
the fifth-dimensional coordinate of anti-de Sitter space $z$ and a
specific impact variable $\zeta$ which measures the separation of
the constituents within the hadron in ordinary space-time. This
connection allows one to predict the form of the  light-front
wavefunctions of mesons and baryons, the fundamental entities which
encode hadron properties and scattering amplitudes. A new
relativistic Schr\"odinger light-front equation is found which
reproduces the results obtained using the fifth-dimensional theory. Since they are complete and orthonormal, the AdS/CFT model
wavefunctions can be used as an initial ansatz for a variational
treatment or as a basis for the diagonalization of the light-front
QCD Hamiltonian. A number of applications of light-front wavefunctions are also discussed.
\end{abstract}

\maketitle
\thispagestyle{fancy}

\section{Hadron Wavefunctions in QCD}

One of the most important tools in atomic physics is the Schr\"odinger
wavefunction; it  provides a quantum mechanical description of the
position and spin coordinates of nonrelativistic bound states at a
given time $t$. Clearly, it is  an important goal in hadron and
nuclear physics  to determine the wavefunctions of hadrons in terms
of their fundamental quark and gluon constituents.

Guy de T\'eramond and I have recently shown how one can use AdS/CFT to not only obtain an accurate
description of the hadron spectrum for light quarks, but also how to obtain a remarkably simple
but realistic model of the valence wavefunctions of mesons, baryons, and glueballs.
As I review below,  the  amplitude $\Phi(z)$ describing  the
hadronic state in the fifth dimension of Anti-de Sitter space
$\rm{AdS}_5$ can be precisely mapped to the
light-front wavefunctions $\psi_{n/h}$ of hadrons in physical
space-time~\cite{Brodsky:2006uq},  thus providing a relativistic
description of hadrons in QCD at the amplitude level.  The
light-front wavefunctions are relativistic and frame-independent
generalizations of the familiar Schr\"odinger wavefunctions of
atomic physics, but they are determined at fixed light-cone time
$\tau= t +z/c$---the ``front form" advocated by Dirac---rather than
at fixed ordinary time $t$.

Formally, the light-front expansion is constructed by quantizing QCD
at fixed light-cone time \cite{Dirac:1949cp} $\tau = t + z/c$ and
forming the invariant light-front Hamiltonian: $ H^{QCD}_{LF} = P^+
P^- - {\vec P_\perp}^2$ where $P^\pm = P^0 \pm
P^z$.~\cite{Brodsky:1997de}   The momentum generators $P^+$ and
$\vec P_\perp$ are kinematical; {\em i.e.}, they are independent of
the interactions. The generator $P^- = i {d\over d\tau}$ generates
light-front time translations, and the eigen-spectrum of the Lorentz
scalar $ H^{QCD}_{LF}$ gives the mass spectrum of the color-singlet
hadron states in QCD together with their respective light-front
wavefunctions.  For example, the proton state satisfies:
$H^{QCD}_{LF} \ket{\psi_p} = M^2_p \ket{\psi_p}$.\linebreak
Remarkably, the light-front wavefunctions are frame-\linebreak
independent;thus knowing the LFWFs of a hadron in its rest frame
determines the wavefunctions in all other frames.

Given thelight-front wavefunctions$\psi_{n/H}(x_i, \vec k_{\perp i},
\lambda_i )$,one can compute a large range of hadron observables.
For example, the valence and sea quark and gluon distributions which
are measured in deep inelastic lepton scattering are defined from
the squares of the LFWFS summed over all Fock states $n$. Form
factors, exclusive weak transition amplitudes~\cite{Brodsky:1998hn}
such as $B\to \ell \nu \pi$. and the generalized parton
distributions~\cite{Brodsky:2000xy}measured in deeply virtual
Compton scattering are (assuming the``handbag" approximation)
overlaps of the initial and final LFWFS with $n =n^\prime$ and $n
=n^\prime+2$. The gauge-invariant distribution amplitude
$\phi_H(x_i,Q)$defined from the integral over the transverse momenta
$\vec k^2_{\perp i} \le Q^2$ of the valence (smallest $n$) Fock
state provides a fundamental measure of the hadron at the amplitude
level~\cite{Lepage:1979zb,Efremov:1979qk}; they  are the
nonperturbative input to the factorized form of hard exclusive
amplitudes and exclusive heavy hadron decays in perturbative QCD.
The resulting distributions obey the DGLAP and ERBL evolution
equations as a function of the maximal invariant mass, thus
providing a physical factorization scheme~\cite{Lepage:1980fj}. In
each case, the derived quantities satisfy the appropriate operator
product expansions, sum rules, and evolution equations. However, at
large $x$ where the struck quark is far-off shell, DGLAP evolution
is quenched~\cite{Brodsky:1979qm}, so that the fall-off of the DIS
cross sections in $Q^2$ satisfies inclusive-exclusive duality at
fixed $W^2.$

The physics of higher Fock states such as the $\vert uud q \bar Q
\rangle$ fluctuation of the proton is nontrivial, leading to
asymmetric $s(x)$ and $\bar s(x)$ distributions, $\bar u(x) \ne \bar
d(x)$, and intrinsic heavy quarks $c \bar c$ and $b \bar b$ which
have their support at high momentum~\cite{Brodsky:2000sk}. Color
adds an extra element of complexity: for example there are
five-different color singlet combinations of six $3_C$ quark
representations which appear in the deuteron's valence wavefunction,
leading to the hidden color phenomena~\cite{Brodsky:1983vf}.

\begin{figure}[htb]
\centering
\includegraphics{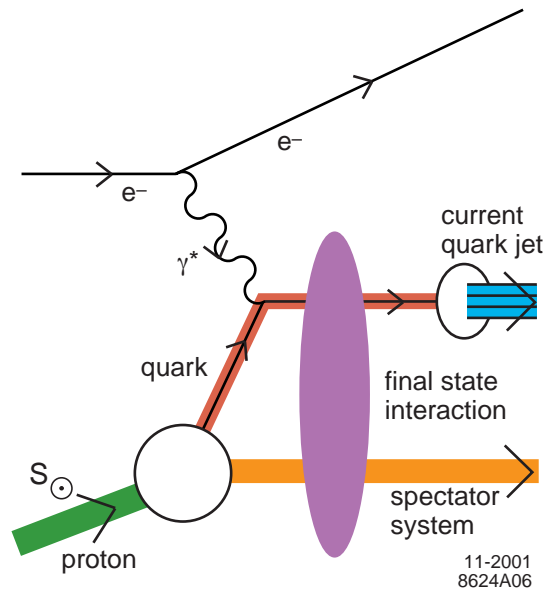}
\vspace{1.5cm}       
\caption{A final-state interaction from gluon exchange in deep
inelastic lepton scattering. The difference of the QCD
Coulomb-like  phases in different orbital states of the proton
produces a single proton spin asymmetry.} \label{SSA}
\end{figure}

An important example of the utility of light-front wavefunctions in
hadron physics is the computation of polarization effects  such as
the single-spin azimuthal asymmetries  in semi-inclusive deep
inelastic scattering, representing the correlation of the spin of
the proton target and the virtual photon to hadron production plane:
$\vec S_p \cdot \vec q \times \vec p_H$. Such asymmetries are
time-reversal odd, but they can arise in QCD through phase
differences in different spin amplitudes. In fact, final-state
interactions from gluon exchange between the outgoing quarks and the
target spectator system lead to single-spin asymmetries in
semi-inclusive deep inelastic lepton-proton scattering  which  are
not power-law suppressed at large photon virtuality $Q^2$ at fixed
$x_{bj}$~\cite{Brodsky:2002cx} (see: Fig.~\ref{SSA}). In contrast to
the SSAs arising from transversity and the Collins fragmentation
function, the fragmentation of the quark into hadrons is not
necessary; one predicts a correlation with the production plane of
the quark jet itself. Physically, the final-state interaction phase
arises as the infrared-finite difference of QCD Coulomb phases for
hadron wave functions with differing orbital angular momentum.  The
same proton matrix element which determines the spin-orbit
correlation $\vec S \cdot \vec L$ also produces the anomalous
magnetic moment of the proton, the Pauli form factor, and the
generalized parton distribution $E$ which is measured in deeply
virtual Compton scattering. Thus the contribution of each quark
current to the SSA is proportional to the contribution
$\kappa_{q/p}$ of that quark to the proton target's anomalous
magnetic moment $\kappa_p = \sum_q e_q
\kappa_{q/p}$.~\cite{Brodsky:2002cx,Burkardt:2004vm}. The HERMES
collaboration has recently measured the SSA in pion
electroproduction using transverse target
polarization~\cite{Airapetian:2004tw}. The Sivers and Collins
effects can be separated using planar correlations; both processes
are observed to contribute, with values not in disagreement with
theory expectations~\cite{Airapetian:2004tw,Avakian:2004qt}. The
deeply virtual Compton amplitudes can be Fourier transformed to
$b_\perp$ and $\sigma = x^-P^+/2$ space providing new insights into
QCD
distributions~\cite{Burkardt:2005td,Ji:2003ak,Brodsky:2006in,Hoyer:2006xg}.
The distributions in the LF direction $\sigma$ typically display
diffraction patterns arising from the interference of the initial
and final state LFWFs~\cite{Brodsky:2006in}.
 The final-state interaction mechanism provides an appealing physical
explanation within QCD of single-spin asymmetries. Physically, the
final-state interaction phase arises as the infrared-finite
difference of QCD Coulomb phases for hadron wave functions with
differing orbital angular momentum.  An elegant discussion of the
Sivers effect including its sign has been given by
Burkardt~\cite{Burkardt:2004vm}. As shown by Gardner and
myself~\cite{Brodsky:2006ha}, one can also use the Sivers effect to
study the orbital angular momentum of  gluons by tagging a gluon jet
in semi-inclusive DIS. In this case, the final-state interactions
are enhanced by the large color charge of the gluons.

The final-state interaction effects can also be identified with the
gauge link which is present in the gauge-invariant definition of
parton distributions~\cite{Collins:2004nx}.  Even when the
light-cone gauge is chosen, a transverse gauge link is required.
Thus in any gauge the parton amplitudes need to be augmented by an
additional eikonal factor incorporating the final-state interaction
and its phase~\cite{Ji:2002aa,Belitsky:2002sm}. The net effect is
that it is possible to define transverse momentum dependent parton
distribution functions which contain the effect of the QCD
final-state interactions.

A related analysis also predicts that the initial-state interactions
from gluon exchange between the incoming quark and the target
spectator system lead to leading-twist single-spin asymmetries in
the Drell-Yan process\linebreak
 $H_1 H_2^\updownarrow \to \ell^+ \ell^- X$
\cite{Collins:2002kn,Brodsky:2002rv}.    Initial-state interactions also lead
to a $\cos 2 \phi$ planar correlation in unpolarized Drell-Yan
reactions \cite{Boer:2002ju}.

\section{Diffractive Deep Inelastic Scattering}

A remarkable feature of deep inelastic lepton-proton scattering at
HERA is that approximately 10\% events are
diffractive~\cite{Adloff:1997sc,Breitweg:1998gc}: the target proton
remains intact, and there is a large rapidity gap between the proton
and the other hadrons in the final state.  These diffractive deep
inelastic scattering (DDIS) events can be understood most simply
from the perspective of the color-dipole model: the $q \bar q$ Fock
state of the high-energy virtual photon diffractively dissociates
into a diffractive dijet system.  The exchange of multiple gluons
between  the color dipole of the $q \bar q$ and the quarks of the
target proton neutralizes the color separation and leads to the
diffractive final state.  The same multiple gluon exchange also
controls diffractive vector meson electroproduction at large photon
virtuality~\cite{Brodsky:1994kf}.  This observation presents a
paradox: if one chooses the conventional parton model frame where
the photon light-front momentum is negative $q+ = q^0 + q^z  < 0$,
the virtual photon interacts with a quark constituent with
light-cone momentum fraction $x = {k^+/p^+} = x_{bj}$.  Furthermore,
the gauge link associated with the struck quark (the  Wilson line)
becomes unity in light-cone gauge $A^+=0$. Thus the struck
``current" quark apparently experiences no final-state interactions.
Since the light-front wavefunctions $\psi_n(x_i,k_{\perp i})$ of a
stable hadron are real, it appears impossible to generate the
required imaginary phase associated with pomeron exchange, let alone
large rapidity gaps.

This paradox was resolved by Paul Hoyer, Nils Marchal, Stephane
Peigne, Francesco Sannino and myself~\cite{Brodsky:2002ue}. Consider
the case where the virtual photon interacts with a strange
quark---the $s \bar s$ pair is assumed to be produced in the target
by gluon splitting.  In the case of Feynman gauge, the struck $s$
quark continues to interact in the final state via gluon exchange as
described by the Wilson line. The final-state interactions occur at
a light-cone time $\Delta\tau \simeq 1/\nu$ shortly after the
virtual photon interacts with the struck quark. When one integrates
over the nearly-on-shell intermediate state, the amplitude acquires
an imaginary part. Thus the rescattering of the quark produces a
separated color-singlet $s \bar s$ and an imaginary phase. In the
case of the light-cone gauge $A^+ = \eta \cdot A =0$, one must also
consider the final-state interactions of the (unstruck) $\bar s$
quark. The gluon propagator  in light-cone gauge $d_{LC}^{\mu\nu}(k)
= (i/k^2+ i \epsilon)\left[-g^{\mu\nu}+\left(\eta^\mu k^\nu+
k^\mu\eta^\nu / \eta\cdot k\right)\right] $ is singular at $k^+ =
\eta\cdot k = 0$. The momentum of the exchanged gluon $k^+$ is of $
\mathcal{ O}{(1/\nu)}$; thus rescattering contributes at leading
twist even in light-cone gauge. The net result is  gauge invariant
and is identical to the color dipole model calculation. The
calculation of the rescattering effects on DIS in Feynman and
light-cone gauge through three loops is given in detail for an
Abelian model in the references~\cite{Brodsky:2002ue}.    The result
shows that the rescattering corrections reduce the magnitude of the
DIS cross section in analogy to nuclear shadowing.

A new understanding of the role of final-state interactions in deep
inelastic scattering has thus emerged. The multiple  scattering of
the struck parton via instantaneous interactions in the target
generates dominantly imaginary diffractive amplitudes, giving rise
to an effective ``hard pomeron" exchange.  The presence of a
rapidity gap between the target and diffractive system requires that
the target remnant emerges in a color-singlet state; this is made
possible in any gauge by the soft rescattering. The resulting
diffractive contributions leave the target intact  and do not
resolve its quark structure; thus there are contributions to the DIS
structure functions which cannot be interpreted as parton
probabilities~\cite{Brodsky:2002ue}; the leading-twist contribution
to DIS  from rescattering of a quark in the target is a coherent
effect which is not included in the light-front wave functions
computed in isolation. One can augment the light-front wave
functions with a gauge link corresponding to an external field
created by the virtual photon $q \bar q$ pair
current~\cite{Belitsky:2002sm,Collins:2004nx}.   Such a gauge link
is process dependent~\cite{Collins:2002kn}, so the resulting
augmented LFWFs are not
universal~\cite{Belitsky:2002sm,Brodsky:2002ue,Collins:2003fm}. We
also note that the shadowing of nuclear structure functions is due
to the destructive interference between multi-nucleon amplitudes
involving diffractive DIS and on-shell intermediate states with a
complex phase. In contrast, the wave function of a stable target is
strictly real since it does not have on-energy-shell intermediate
state configurations. The physics of rescattering and shadowing is
thus not included in the nuclear light-front wave functions, and a
probabilistic interpretation of the nuclear DIS cross section is
precluded.

Rikard Enberg, Paul Hoyer, Gunnar Ingelman and
I~\cite{Brodsky:2004hi} have shown that the quark structure function
of the effective hard pomeron has the same form as the quark
contribution of the gluon structure function. The hard pomeron is
not an intrinsic part of the proton; rather it must be considered as
a dynamical effect of the lepton-proton interaction. Our QCD-based
picture also applies to diffraction in hadron-initiated processes.
The rescattering is different in virtual photon- and hadron-induced
processes due to the different color environment, which accounts for
the  observed non-universality of diffractive parton distributions.
This framework also provides a theoretical basis for the
phenomenologically successful Soft Color Interaction(SCI)
model~\cite{Edin:1995gi}which includes rescattering effects and thus
generates a variety of final states with rapidity gaps.

The phase structure of hadron matrix elements is thus an essential
feature of hadron dynamics. Although the LFWFs are real for a stable
hadron, they acquire phases from initial state and final state
interactions.   In addition, the violation of $CP$ invariance leads
to a specific phase structure of the LFWFs.  Dae Sung Hwang, Susan
Gardner and I~\cite{Brodsky:2006ez} have shown that this in turn
leads to the electric dipole moment of the hadron and a general
relation between the edm and anomalous magnetic moment Fock state by
Fock state.

There are also leading-twist diffractive contributions  $\gamma^*
N_1 \to (q \bar q) N_1$  arising from Reggeon exchanges in the
$t$-channel~\cite{Brodsky:1989qz}.  For example,
isospin--non-singlet $C=+$ Reggeons contribute to the difference of
proton and neutron structure functions, giving the characteristic
Kuti-Weisskopf $F_{2p} - F_{2n} \sim x^{1-\alpha_R(0)} \sim x^{0.5}$
behavior at small $x$. The $x$ dependence of the structure functions
reflects the Regge behavior $\nu^{\alpha_R(0)} $ of the virtual
Compton amplitude at fixed $Q^2$ and $t=0$. The phase of the
diffractive amplitude is determined by analyticity and crossing to
be proportional to $-1+ i$ for $\alpha_R=0.5,$ which together with
the phase from the Glauber cut, leads to {\it constructive}
interference of the diffractive and nondiffractive multi-step
nuclear amplitudes. Furthermore, because of its $x$ dependence, the
nuclear structure function is enhanced precisely in the domain $0.1
< x <0.2$ where antishadowing is empirically observed.  The strength
of the Reggeon amplitudes is fixed by the fits to the nucleon
structure functions, so there is little model dependence.
 Ivan Schmidt, Jian-Jun Yang, and
I~\cite{Brodsky:2004qa} have applied this analysis to the shadowing
and antishadowing of all of the electroweak structure functions.
Quarks of different flavors  will couple to different Reggeons; this
leads to the remarkable prediction that nuclear antishadowing is not
universal; it depends on the quantum numbers of the struck quark.
This picture leads to substantially different antishadowing for
charged and neutral current reactions, thus affecting the extraction
of the weak-mixing angle $\theta_W$.  We find that part of the
anomalous NuTeV result~\cite{Zeller:2001hh} for $\theta_W$ could be
due to the non-universality of nuclear antishadowing for charged and
neutral currents. Detailed measurements of the nuclear dependence of
individual quark structure functions are thus needed to establish
the distinctive phenomenology of shadowing and antishadowing and to
make the NuTeV results definitive.   Antishadowing can also depend
on the target and beam polarization.

\section{The Conformal Approximation to QCD}

One of the most interesting recent developments in hadron physics
has been the use of  Anti-de Sitter space holographic methods in
order to obtain a first approximation to nonperturbative QCD. The
essential principle underlying the AdS/CFT approach to conformal
gauge theories is the isomorphism of the group of Poincare' and
conformal transformations $SO(2,4)$ to the group of isometries of
Anti-de Sitter space.  The AdS metric is
\[ds^2 = {R^2\over z^2}(\eta^{\mu \nu} dx_\mu
dx^\mu - dz^2),\]which is invariant under scale changes of the
coordinate in the fifth dimension $z \to \lambda z$ and $ dx_\mu \to
\lambda dx_\mu$. Thus one can match scale transformations of the
theory in $3+1$ physical space-time to scale transformations in the
fifth dimension $z$. The amplitude $\phi(z)$ represents the
extension of the hadron into the fifth dimension.  The behavior of
$\phi(z) \to z^\Delta$ at $z \to 0$ must match the twist-dimension
of the hadron at short distances $x^2 \to 0$.   As shown by
Polchinski and Strassler~\cite{Polchinski:2001tt}, one can simulate
confinement by imposing the condition $\phi(z = z_0 ={1\over
\Lambda_{QCD}})$.  This approach, has been successful in reproducing
general properties of scattering processes of QCD bound
states~\cite{Polchinski:2001tt,Brodsky:2003px}, the low-lying hadron
spectra~\cite{deTeramond:2005su,Erlich:2005qh}, hadron couplings and
chiral symmetry breaking~\cite{Erlich:2005qh,Hong:2005np}, quark
potentials in confining backgrounds~\cite{Boschi-Filho:2005mw} and
pomeron physics~\cite{Brower-Polchisnki}.

It was originally believed that the AdS/CFT mathematical tool could
only be applicable to strictly conformal theories such as
$\mathcal{N}=4$ supersymmetry. However, if one considers a
semi-classical approximation to QCD with massless quarks and without
particle creation or absorption, then the resulting $\beta$ function
is zero, the coupling is constant, and the approximate theory is
scale and conformal invariant. Conformal symmetry is of course
broken in physical QCD; nevertheless, one can use conformal symmetry
as a {\it template}, systematically correcting for its nonzero
$\beta$ function as well as higher-twist effects. For example,
``commensurate scale relations"~\cite{Brodsky:1994eh}which relate
QCD observables to each other, such as the generalized Crewther
relation~\cite{Brodsky:1995tb}, have no renormalization scale or
scheme ambiguity and retain a convergent perturbative structure
which reflects the underlying conformal symmetry of the classical
theory.  In general, the scale is set such that one has the correct
analytic behavior at the heavy particle
thresholds~\cite{Brodsky:1982gc}.

In a confining theory where gluons have an effective mass, all
vacuum polarization corrections to the gluon self-energy decouple at
long wavelength.  Theoretical~\cite{Alkofer:2004it} and
phenomenological~\cite{Brodsky:2002nb} evidence is in fact
accumulating that QCD couplings based on physical observables such
as $\tau$ decay~\cite{Brodsky:1998ua} become constant at small
virtuality;{\em i.e.}, effective charges develop an infrared fixed
point in contradiction to the usual assumption of singular growth in
the infrared. The near-constant behavior of effective couplings also
suggests that QCD can be approximated as a conformal theory even at
relatively small momentum transfer. The importance of using an
analytic effective charge~\cite{Brodsky:1998mf} such as the pinch
scheme~\cite{Binger:2006sj,Cornwall:1989gv} for unifying the
electroweak and strong couplings and forces is also
important~\cite{Binger:2003by}.  Thus conformal symmetry is a useful
first approximant even for physical QCD.

\begin{figure*}
\includegraphics{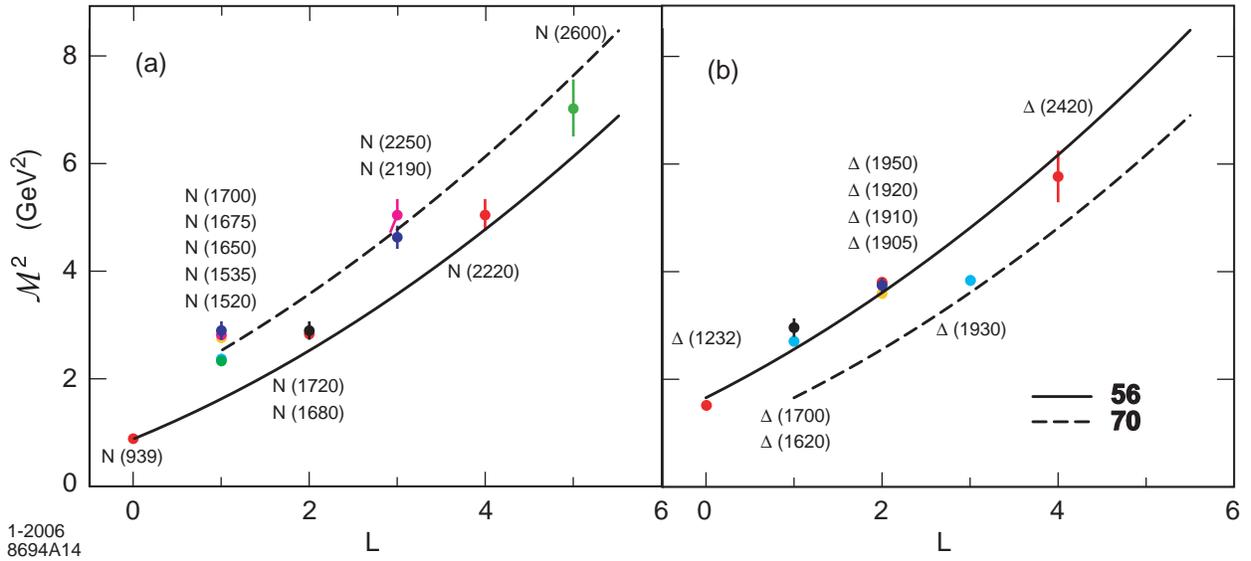}
\vspace{1.5cm}
\caption{Predictions for the light baryon orbital spectrum for
$\Lambda_{QCD}$ = 0.25 GeV. The  $\bf 56$ trajectory corresponds to
$L$ even  $P=+$ states, and the $\bf 70$ to $L$ odd  $P=-$ states.}
\label{fig:BaryonSpec}       
\end{figure*}

\section{Hadronic Spectra in AdS/QCD}

Guy de T\'eramond and I~\cite{Brodsky:2006uq,deTeramond:2005su} have
recently shown how a holographic model based on truncated AdS space
can be used to obtain the hadronic spectrum of light quark $q \bar
q, qqq$ and $gg$ bound states. Specific hadrons are identified by
the correspondence of the amplitude in the fifth dimension with the
twist-dimension of the interpolating operator for the hadron's
valence Fock state, including its orbital angular momentum
excitations. An interesting  aspect of our approach is to show that
the mass parameter $\mu R$ which appears in the string theory in the
fifth dimension is quantized, and that it appears as a Casimir
constant governing the orbital angular momentum of the hadronic
constituents analogous to $L(L+1)$ in the radial Schr\"odinger
equation.

As an example, the set of three-quark baryons with spin 1/2 and
higher  is described  in AdS/CFT by the Dirac equation in the fifth
dimension~\cite{Brodsky:2006uq}
\begin{equation}
\left[z^2~ \partial_z^2 - 3 z~ \partial_z + z^2 \mathcal{M}^2 -
\mathcal{L}_\pm^2 + 4\right] \psi_\pm(z) = 0.
\end{equation}
The constants $\mathcal{L}_+  = L + 1$, $\mathcal{L}_- = L + 2$ in
this equation are  Casimir constants which are determined to match
the twist dimension of the solutions with arbitrary relative orbital
angular momentum. The solution is\begin{equation}
\label{eq:DiracAdS} \Psi(x,z) = C e^{-i P \cdot x} \left[\psi(z)_+~
u_+(P) + \psi(z)_-~ u_-(P) \right],
\end{equation}with $\psi_+(z) = z^2 J_{1+L}(z \mathcal{M})$ and $\psi_-(z) = z^2
J_{2+L}(z \mathcal{M})$. The physical string solutions have plane
waves and chiral spinors $u(P)_\pm$ along the Poincar\'e coordinates
and hadronic invariant mass states given by $P_\mu P^\mu =
\mathcal{M}^2$. A discrete  four-dimensional spectrum follows when
we impose the boundary condition $\psi_\pm(z=1/\Lambda_{\rm QCD}) =
0$. One has $\mathcal{M}_{\alpha, k}^+ = \beta_{\alpha,k}
\Lambda_{\rm QCD},$ $\mathcal{M}_{\alpha, k}^- = \beta_{\alpha +
1,k} \Lambda_{\rm QCD}$, with a scale-independent mass
ratio~\cite{deTeramond:2005su}. The $\beta_{\alpha,k}$ are the first
zeros of the Bessel eigenfunctions.

Figure \ref{fig:BaryonSpec}(a) shows the predicted orbital spectrum
of the nucleon states and Fig.~\ref{fig:BaryonSpec}(b) the $\Delta$
orbital resonances. The spin 3/2 trajectories are determined from
the corresponding Rarita-Schwinger  equation. The data for the
baryon spectra are from S. Eidelman {\em et
al.}~\cite{Eidelman:2004wy}. The internal parity of states is
determined from the SU(6) spin-flavor symmetry.

Since only one parameter, the QCD mass scale $\Lambda_{QCD}$, is
introduced, the agreement with the pattern of physical states is
remarkable. In particular, the ratio of $\Delta$ to nucleon
trajectories is determined by the ratio of zeros of Bessel
functions.  The predicted mass spectrum in the truncated space model
is linear $M \propto L$ at high orbital angular momentum, in
contrast to the quadratic dependence $M^2 \propto L$ in the usual
Regge parametrization.

\begin{figure}
\resizebox{0.50\textwidth}{!}{%
  \includegraphics{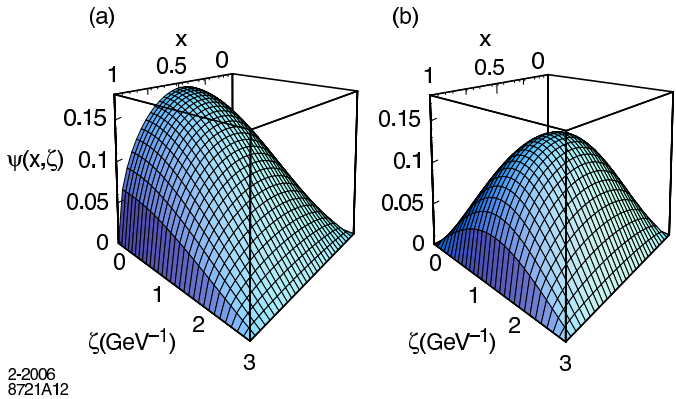}
}
\caption{AdS/QCD Predictions for the $L=0$ and $L=1$ LFWFs of a
meson}
\label{fig:MesonLFWF}       
\end{figure}

Our approach shows that  there is an exact correspondence between
the fifth-dimensional coordinate of anti-de Sitter space $z$ and a
specific impact variable $\zeta$ in the light-front formalism which
measures the separation of the constituents within the hadron in
ordinary space-time.   The  amplitude $\Phi(z)$ describing  the
hadronic state in $\rm{AdS}_5$ can be precisely mapped to the
light-front wavefunctions $\psi_{n/h}$ of hadrons in physical
space-time\cite{Brodsky:2006uq},  thus providing a relativistic
description of hadrons in QCD at the amplitude level. We derived
this correspondence by noticing that the mapping of $z \to \zeta$
analytically transforms the expression for the form factors in
AdS/CFT to the exact Drell-Yan-West expression in terms of
light-front wavefunctions. In the case of a two-parton constituent
bound state the correspondence between the string amplitude
$\Phi(z)$ and the light-front wave function
$\widetilde\psi(x,\mathbf{b})$ is expressed in closed
form~\cite{Brodsky:2006uq}
\begin{equation}  \label{eq:Phipsi}
\left\vert\widetilde\psi(x,\zeta)\right\vert^2 = \frac{R^3}{2 \pi}
~x(1-x)~ e^{3 A(\zeta)}~ \frac{\left\vert
\Phi(\zeta)\right\vert^2}{\zeta^4},
\end{equation}
where $\zeta^2 = x(1-x) \mathbf{b}_\perp^2$. Here $b_\perp$ is the
impact separation and Fourier conjugate to $k_\perp$. The variable
$\zeta$, $0 \le \zeta \le \Lambda^{-1}_{\rm QCD}$, represents the
invariant separation between point-like constituents, and it is also
the holographic variable $z$ in AdS; {\em i.e.}, we can identify
$\zeta = z$. The prediction for the meson light-front wavefunction
is shown in Fig.~\ref{fig:MesonLFWF}. We can also transform the
equation of motion in the fifth dimension using the $z$ to $\zeta$
mapping to obtain an effective two-particle light-front radial
equation
\begin{equation}
\label{eq:Scheq} \left[-\frac{d^2}{d \zeta^2} + V(\zeta) \right]
\phi(\zeta) = \mathcal{M}^2 \phi(\zeta),
\end{equation}
with the effective potential $V(\zeta) \to - (1-4 L^2)/4\zeta^2$ in
the conformal limit. The solution to (\ref{eq:Scheq}) is $\phi(z) =
z^{-\frac{3}{2}} \Phi(z) = C z^\frac{1}{2} J_L(z \mathcal{M})$. This
equation reproduces the AdS/CFT\linebreak
 solutions. The lowest stable state
is determined by the\linebreak
 Breitenlohner-Freedman
bound~\cite{Breitenlohner:1982jf} and its eigenvalues by the boundary
conditions at $\phi(z = 1/\Lambda_{\rm QCD}) = 0$ and given in terms
of the roots of the Bessel functions: $\mathcal{M}_{L,k} =
\beta_{L,k} \Lambda_{\rm QCD}$. Normalized LFWFs follow from
(\ref{eq:Phipsi})
\begin{equation}
\widetilde \psi_{L,k}(x, \zeta) =  B_{L,k} \sqrt{x(1-x)} J_L
\left(\zeta \beta_{L,k} \Lambda_{\rm QCD}\right) \theta\big(z \le
\Lambda^{-1}_{\rm QCD}\big),
\end{equation}
where $B_{L,k} = \pi^{-\frac{1}{2}} {\Lambda_{\rm QCD}} \
J_{1+L}(\beta_{L,k})$. The resulting wavefunctions (see:
Fig.~\ref{fig:MesonLFWF}) display confinement at large inter-quark
separation and conformal symmetry at short distances, reproducing
dimensional counting rules for hard exclusive processes in agreement
with perturbative QCD.

\begin{figure}
\includegraphics{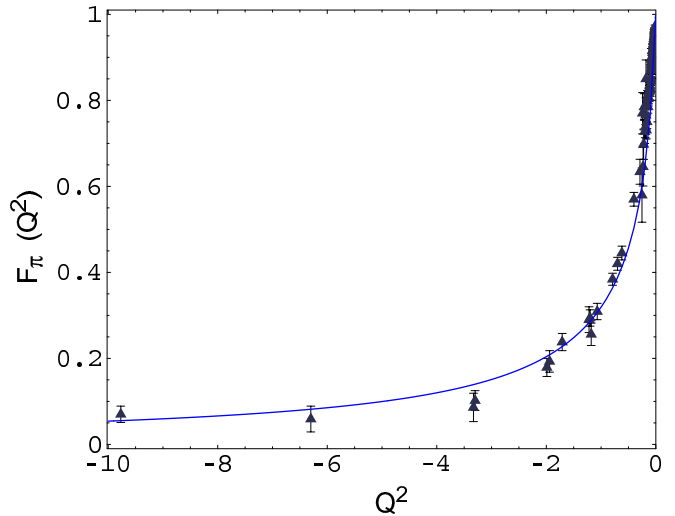}
\vspace{1cm}       
\caption{AdS/QCD Predictions for the pion form factor.}
\label{fig:PionFF}       
\end{figure}

The hadron form factors can be predicted from overlap  integrals in
AdS space or equivalently by using the Drell-Yan-West formula in
physical space-time.  The prediction for the  pion form factor is
shown in Fig.~\ref{fig:PionFF}. The form factor at high $Q^2$
receives contributions from small $\zeta$, corresponding to small
$\vec b_\perp = {\cal O}(1/Q)$ ( high relative $\vec k_\perp = {\cal
O}(Q)$  as well as $x \to 1$.  The AdS/CFT dynamics is thus distinct
from endpoint models~\cite{Radyushkin:2006iz} in which the LFWF is
evaluated solely at small transverse momentum or large impact
separation.

The $x \to 1$ endpoint domain is often referred to as a "soft" Feynman contribution.
In fact $x \to 1$ for the struck quark requires that all of the spectators have $x = k^+/P^+ = (k^0+ k^z)/P^+  \to 0$; this  in turn requires high longitudinal momenta $k^z \to - \infty$ for all spectators  --  unless one has both massless spectator quarks $m \equiv 0$ with zero transverse momentum $k_\perp \equiv 0$, which is a regime of measure zero.
If one uses a covariant formalism, such as the Bethe-Salpeter theory, then the virtuality of the struck quark  becomes  infinitely spacelike:  $k^2_F \sim  - {k^2_\perp + m^2\over 1-x}$  in the endpoint domain.
Thus, actually,  $x \to 1$ corresponds to high relative longitudinal momentum; it is as hard a domain in the hadron wavefunction as high transverse momentum.

It is also interesting to note that the distribution amplitude
predicted by AdS/CFT at the hadronic scale is $\phi_\pi(x, Q ) =
{4\over  \sqrt 3 \pi}  f_\pi \sqrt{x(1-x)}$ from both the harmonic
oscillator and truncated space models is quite different than the
asymptotic distribution amplitude predicted from the PQCD
evolution~\cite{Lepage:1979zb} of the pion distribution amplitude
$\phi_\pi(x,Q \to \infty)= \sqrt 3  f_\pi x(1-x) $.  The broader
shape of the pion distribution increases the magnitude of the
leading twist perturbative QCD prediction for the pion form factor
by a factor of $16/9$ compared to the prediction based on the
asymptotic form, bringing the PQCD prediction  close to the
empirical pion form factor~\cite{Choi:2006ha}.

Since they are complete and orthonormal, the AdS/CFT model
wavefunctions can be used as an initial ansatz for a variational
treatment or as a basis for the diagonalization of the light-front
QCD Hamiltonian.  We are now in fact investigating this possibility
with J. Vary and A. Harindranath. The wavefunctions predicted by
AdS/QCD have many phenomenological applications ranging from
exclusive $B$ and $D$ decays, deeply virtual Compton scattering and
exclusive reactions such as form factors, two-photon processes, and
two body scattering. A connection between the theories and tools
used in string theory and the fundamental constituents of matter,
quarks and gluons, has thus been found.

The application of AdS/CFT to QCD phenomenology is now being
developed in many new directions, incorporating finite quark masses,
chiral symmetry breaking, asymptotic freedom, and finite temperature
effects. Some recent papers are given in refs.
\cite{Evans:2006ea,Kim:2006ut,Csaki:2006ji,Erdmenger:2006bg,Boschi-Filho:2006pt,Evans:2006dj,Harada:2006di,Horowitz:2006ct,Klebanov:2005mh,Shuryak:2006yx}.

\bigskip
\noindent{\bf Acknowledgments}

\vspace{4pt}Work supported by the Department of Energy under
contract number DE--AC02--76SF00515. The AdS/CFT results reported
here were done in collaboration with Guy de T\'eramond.
\bigskip
\bigskip

\end{document}